# SCIENCE-BASED COMPARATIVE CULTURE: A NEW THEME OF EXPERIMENT FOR FRESHMEN IN TOHOKU UNIVERSITY

Tsuyoshi HONDOU[*], Masayuki YOSHIZAWA, and Shozo SUTO
Department of Physics, Tohoku University, Sendai 980-8578 Japan

**ABSTRACT**

In 2004, Tohoku University created a new introductory science experimental course for freshmen [1]. The course is a compulsory subject for students in all natural science fields. The course is not designed for a professional education, but as a liberal education, in which students are trained to become familiar with nature and to discover natural laws for themselves. We present here one of 12 themes - "science and culture: vibration of string instrument and music", in which we expect students to study two aspects: 1) the universality of natural laws and 2) the variety of value judgments from the evidence.

## 1. UNIVERSALITY OF NATURAL LAW

In this experiment, students play classical guitars (Fig. 1). Using the "harmonics (flageolet)" technique, as in Figs. 2 and 3, students intuitively distinguish higher harmonic waves (up to $6^{th}$ harmonics) using their ears and eyes, without special scientific instruments. If the fundamental tone of the string is Do (C), tone names of the higher harmonics are Do (C), Sol (G), Do (C), Mi (E) and Sol (G), which are the origins (bases) of harmony and musical scale ("natural musical scale"). The ratios of frequency Sol/Do, Mi/Do are 3/2 and 5/4, respectively, which are rational numbers. Students will find the physical basis of music that is common over ethnic groups, because of the universality of natural laws.

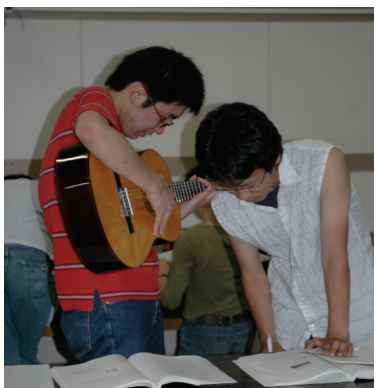

Fig. 1 Discover natural laws by ears and eyes

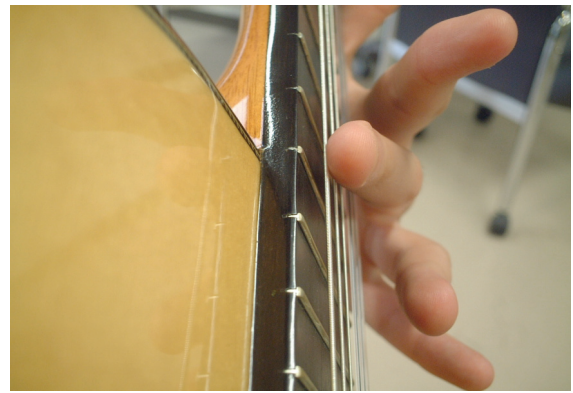

Fig. 2 Mode selection by harmonics method

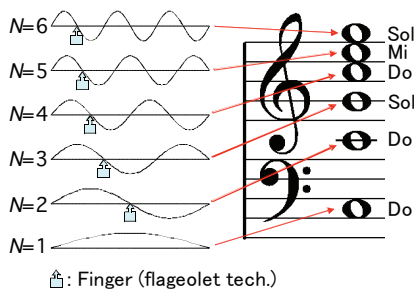

Fig. 3 Modes of string's vibration and musical notes.
Students will find by themselves:
1) Musical scale is quantized corresponding to modes.
2) Musical score is applicable for almost all music, because natural laws are universal, independent of regions, times.
3) So, music is called a "universal language" in which Do, Mi, Sol are always important.

## 2. VARIETY OF VALUE JUDGMENT

In most science experiments, students learn universal aspects of natural laws. In this experiment, students also learn a variety of value judgments through comparing two musical scales - "equal temperament" which was introduced in Europe and the natural musical scale.

Equal temperament is an artificial musical scale devised in Europe, in which 12 notes in 1 octave are "equally" divided. The frequency ratio of the neighboring notes (eg. C# and C) is defined as $2^{1/12}$. Thus, the frequency ratio of two notes is irrational (except for octave), making a "beat" in the sound of coexisting two notes, which is in contrast to the sound of natural musical scale. On the other hand, equal temperament has the freedom of change in key, which is also in contrast to natural musical scale. Thus, we have different musical scales used in the present day. The choice of one musical scale is a result of a "value judgment" which has diversity depending on persons, ethnic groups, period and so on.

In these experiments, students analyze the difference of frequencies between natural musical scale and equal temperament. Students are then asked why equal temperament was devised in Europe, in spite of its weakness that it makes sound impure through beating. Students will find why European people chose equal temperament for piano music, in spite of the negative aspect of this musical scale. This is just a "science-based comparative study".

## 3. SCIENTIFIC LITERACY: EVIDENCE-BASED DECISION

In several cases, confusion between scientific evidence and value judgment is observed, which constricts proper utilization of science in society. With the materials in this experiment, we encourage students to be familiar with the Evidence-Based Decision (EBD), in which Evidence-Based Medicine (EBM) is the most famous. In the experiment, students find a historical example of EBD through the choice of equal temperament. They find that study of natural law is useful to find cultural diversity based on diversity of value judgment.

## 4. FUTURE AND ACKNOWLEGMENT

We are now expanding this experimental course to students in the Faculties of Education, Arts and Letters, Economies and Law (in a modified form), which will start in 2007. We think that "science-based comparative culture" is an ideal material for "physics for all", since evidence-based decision, a key concept in scientific literacy, can be studied by using the five senses. We hope a variety of materials used in this approach will be developed for the future. We especially thank Mr. Inami (Japan Philharmonic Orchestra, trombone) and Mr. Saeki (composer) for their help.

Corresponding Author
Tsuyoshi HONDOU, Department of Physics, Tohoku University, Sendai 980-8578 Japan,
E-mail hondou@cmpt.phys.tohoku.ac.jp